\documentstyle[12pt]{article}

 \font\cero=cmbx10 scaled 1728

\begin{document}
\begin{flushleft}
{\cero N-dimensional geometries and Einstein equations from
systems of PDEs}
\end{flushleft}
Emanuel Gallo$^{\spadesuit}$, Magdalena
Marciano-Melchor$^{\clubsuit}$ and Gilberto
Silva-Ortigoza$^{\clubsuit}$\\[2em]

\noindent$^{\spadesuit}$  {\it FaMAF, Universidad Nacional de
C\'{o}rdoba, 5000 C\'{o}rdoba, Argentina.}\\[1em]

\noindent $^{\clubsuit}${\it Facultad de Ciencias F\'{\i}sico
Matem\'aticas de la Universidad Aut\'onoma de Puebla, Apartado
Postal 1152, Puebla, Pue., M\'exico.}\\[1em]


\begin{abstract}
The aim of the present work is twofold: first, we show how all the
$n$-dimensional Riemannian and Lorentzian metrics can be
constructed from a certain class of systems of second-order PDE's
which are in duality to the Hamilton-Jacobi equation and second we
impose the Einstein equations to these PDE's.
\end{abstract}

\newpage

\section{Introduction}

Towards the end of the 19th century and early 20th century Tresse, W\"{u}ns\-chmann, Lie, Cartan and Chern
(\cite{T1}-\cite{Ch}), studied the classification of second- and third-order ODE's according to their equivalence
classes under a variety of transformations and the resulting induced geometries on the solution spaces. In
particular, Cartan (\cite {C}-\cite{C3}) and Chern \cite{Ch} found that for a certain subclass of third-order
ODE's, a unique Lorentzian conformal metric can be constructed in a natural way on the solution space. This
subclass was defined by the vanishing of a specific relative invariant, defined from the differential equation and
first obtained by W\"{u}nschmann \cite{W}. It is now referred to as the W\"{u}nschmann invariant. In a more recent
work Tod\,\cite{Tod} showed how the Einstein-Weyl spaces can be obtained from a particular class of third-order
differential equations.\\

In a more recent series of papers Frittelli, Kozameh, Newman, Kamran and Nurowski,\thinspace
(\cite{FCN1}-\cite{FNN}) were able to generalize this result. They showed that all four-dimensional conformal
Lorentzian geometries were encoded in equivalence classes (under contact transformations) of pairs of second-order
PDE's that were characterized by the vanishing of an analogous (generalized) W\"{u}nschmann invariant, referred to
as the \textit{metricity conditions}. In this approach, referred to as the null surface formulation of general
relativity, the metric of the space-time is a derived concept. The fundamental objects are two functions,
$Z(x^{a}, s , s^{*})$ and $\Omega (x^{a}, s ,s ^{*})$, of the space-time points $x^{a}$ and parametrized by points
on the sphere; that is, by functions defined on ${\cal M}\times {\cal S}^{2}$ (the sphere bundle over the
space-time). The first of the functions, $Z(x^{a},s ,s ^{*})$, which encodes all the conformal information of the
space-time, describes a sphere's worth of surfaces through each space-time point. It is from these surfaces that a
conformal metric can be constructed. The second function, $\Omega (x^{a},s , s^*)$, which plays the role of a
conformal factor, converts it into any metric in the conformal class. The level surfaces of $ Z(x^{a},s , s^*)$ in
${\cal M}$, for each fixed value of $(s,s^*)$, are null hypersurfaces with respect to this metric. As $(s,s^*)$
take different values on ${\cal S}^{2}$ at a fixed point $x^{a}$ in ${\cal M}$, the normals to the null
hypersurfaces sweep out the
null-cone at $x^{a}$.\\

To establish this new approach to general relativity, these
authors began with a four-dimensional Lorentzian manifold, already
containing a metric $g_{ab}$ and a complete integral to the
eikonal equation

\begin{equation}
g^{ab}(x^{a})\nabla _{a}Z  \nabla _{b}Z=0.  \label{EIK*}
\end{equation}

A complete integral, expressed as,
\begin{equation}
u=Z(x^{a},s ,s ^{*}),  \label{SOL*}
\end{equation}
contains the space-time coordinates, $x^{a},$ and the needed (for
a complete integral) two-parameters $(s ,s ^{*}).$ By constructing
the four functions,
\begin{equation}
\theta ^{i}\equiv (u,\omega ,\omega ^{*},R)\equiv (Z,\partial _{s
}Z,
\partial_{s^*}Z,\partial _{s^* s }Z),  \label{THETA*}
\end{equation}
from Eq.\,(\ref{SOL*}) and its derivatives, and by eliminating
$x^{a},$ via the algebraic inversion
\begin{equation}
x^{a}=X^{a}(s ,s ^{*},\theta ^{i}),  \label{INVERSION*}
\end{equation}
they found that $u=Z(x^{a},s ,s ^{*})$ satisfies in addition to
Eq.\,(\ref{SOL*}) the pair of second-order partial differential
equations in $ s ,$ $s ^{*},$ of the form
\begin{eqnarray}
\partial _{s s }Z & = &\Lambda (Z, \partial
_{s }Z, \partial _{s ^{*}}Z, \partial _{s s
^{*}}Z, s , s ^{*}), \nonumber \\
\partial _{s ^{*}s ^{*}}Z & = &\Lambda ^{*}(Z,
\partial _{s }Z, \partial _{s ^{*}}Z, \partial _{s s ^{*}}Z,
s , s ^{*}).\label{PAIR*}
\end{eqnarray}

The $x^{a},$ in the solution of Eq.\,(\ref{PAIR*}), appear now as
constants of integration. The roles of $x^{a}$ and $(s,$ $s ^{*})$
are thereby interchanged. Note that the metric has disappeared
from the equations.

The question then was, could this procedure be reversed? Could one
start with a pair of equations of the form, (\ref{PAIR*}), and
then find the eikonal equation, (\ref{EIK*}), with a metric
$g^{ab}(x^{a})$?

It was shown that when the functions ($\Lambda, \Lambda ^{*})$
satisfy an integrability condition, a weak inequality and a
certain set of differential conditions (the metricity or
generalized W\"{u}nschmann conditions), the procedure can be
reversed. The solutions to the pair do determine a conformal
four-dimensional Lorentzian metric and, in fact, all conformal
Lorentzian metrics can be obtained from equivalence classes of
equations of the form Eq.\,(\ref{PAIR*}). When certain specific
conditions\,\cite{GKNP,KIP}, in addition to the W\"{u}nschmann
condition, are imposed on the ($\Lambda ,\Lambda ^{*}),$ the
metrics, determined by the solutions, are in the vacuum conformal
Einstein class.\newline

In a recent work we presented a new approach to $4$-dimensional
general relativity, which is similar to the null surface
formulation, but now instead of using a complete integral to the
eikonal equation, we used a complete integral to the
Hamilton-Jacobi equation. The aim of the present work is to
generalize our previous results. \newline

In section 2 we begin with an $n$-dimensional manifold, ${\cal M}$, with no further structure and then investigate
arbitrary ($n-1$)-parameter families of surfaces on ${\cal M}$ given by
\begin{equation}
u=constant=Z(x^{a}, s^i).  \label{Z}
\end{equation}
The $x^{a}$ are local coordinates on ${\cal M}$ and $s^i$ parametrize the families and can take values on an open
neighborhood of a manifold ${\cal N}$ of dimension ($n-1$). More specifically, we then ask when do such families
of surfaces define an $n$-dimensional metric, $g_{ab}(x^{a}) $, such that
\begin{equation}
g^{ab}\nabla _{a}Z(x^{a}, s^{i})\nabla_{b} Z(x^{a}, s^{i})=1.
\label{H-Jint}
\end{equation}
We have here either taken the mass in the H-J equation to be $1$
or put it into the $g^{ab}$ as a factor.\\

By taking ($s^{i}$) derivatives of Eq.\,(\ref{Z}) and eliminating
the $x^{a},$ we will show that the $u = Z(x^{a}, s^{i})$ must also
satisfy a system of $\frac{n(n-1)}{2}$ second-order PDE's
\begin{equation}
\partial_{s^is^j}Z=\Lambda_{ij}(u,w^i,s^i)=\Lambda_{ij}(u,\partial_{s^i}Z,s^i),\label{PDEint}
\end{equation}
where $\Lambda_{ij}$, are restricted to satisfy certain metricity
or `W\"{u}nschmann-like' conditions.\newline

Here $\partial _{s^i}$, denotes the partial derivative with respect to the parameter $s^{i}$. Observe that in the
solutions of Eqs.\,(\ref{PDEint}) $u=Z(x^{a}, s^{i}),$ the $x^{a}$ are $n$ constants of integration for
Eqs.\thinspace(\ref{PDEint}) while the `$s^{i}$' are $n-1$ integration constants for Eq.\thinspace(\ref{H-Jint}).
Observe that $u=Z(x^{a}, s^{i})$ is a complete integral to the Hamilton-Jacobi equation (\ref{H-Jint}).\newline

In this section we also remark that the $n$-dimensional metric
$g_{ab}(x^{a})$ associated with the system of partial differential
equations, (\ref{PDEint}), is invariant under a subset of contact
transformations of the differential equations.\newline

In section 3, we present our new formulation of $n$-dimensional general relativity. For this purpose we substitute
the $n$-dimensional metric already obtained in section 2 into the Einstein equations. From our results we conclude
that the Einstein equations in $n$ dimensions can be reformulated as equations for families of ($n-1$)-dimensional
surfaces given by the level
surfaces of $u=Z(x^{a}, s^{i})$.\\

As in four dimensions, this new point of view can be given in either of two versions. In the first version, the
variables are the $\frac{n(n-1)}{2}$ functions, $\Lambda_{ij}$, of the $2n-1$ variables ($u, w^i, s^{i}),$ i.e.,
the right-side of Eqs.\,(\ref{PDEint}). These functions must satisfy three sets of equations; the integrability
conditions, the W\"{u}nschmann-like conditions and a further condition obtained from the Einstein equations. The
metric, on a $n$-manifold, can be written down directly in terms of these $\frac{n(n-1)}{2}$ functions and their
derivatives. \textit{There is no need to use the set}, Eqs.\,(\ref{PDEint}). In the second version one uses the
same set of $\Lambda_{ij}$ in the right-side of Eqs.\,(\ref{PDEint}) and solves for the $Z(x^{a}, s^{i}).$ The
metric is then written in terms of the $Z(x^{a}, s^{i})$ and its derivatives. The advantage of the first version
is that one does not need to solve Eqs.\,(\ref{PDEint}), but one has to extract (algebraically) the $n$-manifold
from the ($u, \partial_i u)$ while in the second version the four-manifold is explicitly given by the four
constants of integration, $x^{a}.$\\

It is important to remark that no claim is made that this approach to the $n$-dimensional Einstein equations has
any obvious advantage over the usual metric approach. It however does give certain mathematical
insights into the differential geometry associated with general relativity in $n$ dimensions.\\

\section{N-D metrics and the metricity or W\"{u}ns\-chmann-like
conditions}

We start with an $n$-dimensional manifold $\mathcal{M}$ (with local coordinates $ x^{a}=(x^{0},..., x^{n-1}))$ and
assume we are given an $(n-1)$-parameter set of functions $u=Z(x^{a}, s^i)$. As we said the parameters $s^i$ can
take values on an open neighborhood of a manifold ${\cal N}$ of dimension ($n-1$). We also assume that for fixed
values of the parameters $s^{i}$ the level surfaces
\begin{equation}
u=constant=Z(x^{a}, s^{i}),  \label{sol3*}
\end{equation}
locally foliate the manifold ${\cal M}$ and that $u=Z(x^{a},
s^{i})$ satisfies the H-J equation
\begin{equation}
g^{ab}(x^{a})\nabla _{a}Z(x^{a}, s^{i}) \nabla _{b} Z(x^{a},
s^{i})=1,  \label{HJ3}
\end{equation}
for some unknown metric $g_{ab}(x^{a})$.\newline

The basic idea now is to solve Eq.\thinspace(\ref{HJ3}) for the
components of the metric in terms of $\nabla_{a}Z(x^{a}, s^{i})$.
To do so, we will consider a number of parameter derivatives of
the condition (\ref{HJ3}), and then by manipulation of these
derivatives, obtain both the $n$-dimensional metric and the system
of partial differential equations defining the surfaces plus the
conditions these PDE's must satisfy. They will be referred to as
the metricity or W\"{u}nschmann-like conditions.\newline

\textbf{Remark 1}: \textit{The notation is as follows: there will
be two types of differentiation, one is with respect to the local
coordinates, $ x^{a}$, of the manifold ${\cal M}$, denoted by
$\nabla_{a}$ or ``comma a,'' the other is with respect to the
parameters $s^i$ denoted by $\partial_{s^i} \equiv
\partial_{i}$}.\newline

From the assumed existence of $u = Z(x^{a}, s^{i})$, we define $n$ parameterized scalars $\theta ^{A}$ in the
following way
\begin{equation}
\theta ^{A}  = (Z, w^i) \equiv (Z, \partial_i Z). \label{uvwR}
\end{equation}

\textbf{Remark 2:} \textit{For each value of $s^{i}$,
Eqs.\thinspace( \ref{uvwR}) can be thought of as a coordinate
transformation between the $x^{a}$'s and ($u$, $w^i$)}.\newline

We also define the following $\frac{n(n-1)}{2}$ important scalars
\begin{equation}
\tilde{\Lambda}_{ij} = \partial _{ij}Z(x^{a}, s^{i}).
\label{GPUFPO*}
\end{equation}
In what follows we will assume that Eqs.\thinspace (\ref{uvwR}) can be
inverted, i.e., solved for the $x^{a}$'s;
\[
x^{a}=X^{a}(u, w^{i}, s^{i}).
\]
Eqs.\thinspace(\ref{GPUFPO*}) can then be rewritten as
\begin{equation}
\partial _{ij}Z = \Lambda_{ij}(u, w^i, s^{i}).
\label{PDEIIIc}
\end{equation}
This means that the ($n-1$)-parameter family of level surfaces,
Eq.\thinspace(\ref{sol3*}), can be obtained as solutions to the
system of $\frac{n(n-1)}{2}$ second-order PDE's (\ref{PDEIIIc}).
Note that $\Lambda_{ij}$ satisfy the integrability conditions
\begin{equation}
D_{s^k}\Lambda_{ij}=D_{s^i}\Lambda_{kj}=D_{s^j}\Lambda_{ki},\label{IC}
\end{equation}
where

\textbf{Definition 1:} \textit{The total $s^{i}$ derivative of a
function $F = F(u, w^{i}, s^{i})$ is defined by}
\begin{equation}
D_{s^k}F=F_{s^k}+ F_{w^l}\Lambda_{l k}.
\end{equation}

The solution space of Eqs.\thinspace(\ref{PDEIIIc}) is
$n$-dimensional. This can be seen in the following way. The system
of PDE's (\ref{PDEIIIc}) is equivalent to the vanishing of the $n$
one-forms, $\omega ^{A}= (\omega^0, \omega^i)$
\begin{eqnarray}
\omega^0 &\equiv& du -  w^{l} ds^{l}, \nonumber
\\
\omega^i &\equiv& dw^{i}- \Lambda_{im} ds^{m}.
\end{eqnarray}
A simple calculation, using the integrability conditions on $\Lambda_{ij}$, leads to $d\omega ^{A}=0\,(modulo$
$\omega ^{A})$ from which, via the Frobenius Theorem,  we conclude that the solution space of
Eqs.\thinspace(\ref{PDEIIIc}) is $n$-dimensional.\newline

From the $n$ scalars, $\theta ^{A}$, we have their associated gradient basis $\theta ^{A}{}_{,\,a}$ given by
\begin{equation}
\theta ^{A}{}_{,\,a}=\nabla _{a}\theta ^{A}=\{Z_{,\,a},
w^i_{,a}\}, \label{grad3}
\end{equation}
and its dual vector basis $\theta _{A}\,^{a}$, so that
\begin{equation}
\theta _{A}\,^{a}\theta ^{B}{}_{,\,a}=\delta
_{A}\,^{B},\,\,\,\,\theta _{A}\,^{a}\theta ^{A}{}_{,b}=\delta
_{b}\,^{a}.  \label{vdv3}
\end{equation}

It is easier to search for the components of the $n$-dimensional
metric in the gradient basis rather than in the original
coordinate basis. Furthermore, it is preferable to use the
contravariant components rather than the covariant components of
the metric; i.e., we want to determine
\begin{equation}
g^{AB}(x^{a}, s^{i}) = g^{ab}(x^{a})\theta ^{A}{}_{,\,a}\theta
_{,\,b}^{B}.  \label{gijIII}
\end{equation}
The metric components and the W\"{u}nschmann-like conditions are
obtained by repeatedly operating with $\partial _i$ on
Eq.\thinspace(\ref{HJ3}), which, by definition, is
\begin{equation}
g^{00}=g^{ab}Z_{,\,a}Z_{,\,b}=1.  \label{h-j3zeta}
\end{equation}
Applying $\partial _{i}$ to Eq.\thinspace(\ref{h-j3zeta}) yields
$\partial _{i}{g^{00}}=2g^{ab}\partial _{i}Z,_{a}Z,_{b}=0,$ i.e.,
\begin{equation}
g^{i0}=0.  \label{gi0}
\end{equation}

A direct computation shows that
\begin{eqnarray}
\partial_{ji}(g^{00}/2) &=&g^{ab}\partial_{ji}Z,_{a}Z,_{b} + g^{ab}\partial_{i}
Z,_{a}\partial _{j}Z,_{b}  \nonumber \\
&=&g^{ab}\Lambda_{ij, a} Z,_{b}+ g^{ij}=0.  \label{gij}
\end{eqnarray}
Since, by the assumed linear independence of $(Z,_{a},\partial
_{i}Z,_{a}),$
\begin{equation}
\Lambda_{ij, a} = \Lambda _{u}Z,_{a} + \Lambda_{ij, w^k}\partial
_{k}Z,_{a}, \label{Lambda}
\end{equation}
Eq.\thinspace(\ref{gij}), using
Eqs.\thinspace(\ref{h-j3zeta})-(\ref{Lambda}), is equivalent to
\begin{equation}
g^{ij}=-\Lambda_{ij,u}.
\end{equation}

Therefore, the final result is
\begin{equation}
(g^{AB})=\left(
\begin{array}{cc}
1 & 0 \\
0 & -\Lambda_{ij, u}
\end{array}
\right) .  \label{gAB}
\end{equation}

\textbf{Remark 3:} \textit{We require that $\det (g^{ij})= \Delta$ be
different from zero, with}
\begin{equation}
\Delta \equiv  \det(-\Lambda_{ij, u}). \label{delta}
\end{equation}

Finally the metricity or W\"{u}nschmann-like conditions are
obtained from the third derivatives, i.e., from $\partial
_{lji}g^{00}=0$. By a direct computation we obtain that
\begin{equation}
D_{s^k}\left [\Lambda_{mn,u}\right ]=
\Lambda_{ln,u}\Lambda_{km,w^l} +
\Lambda_{lm,u}\Lambda_{kn,w^l}.\label{WLC3D'}
\end{equation}

In $n$ dimensions, with $n \geq 2 $, there will be $\frac{n(n^2-1)}{6}$ W\"unschmann-type conditions. For example
for $n=2$ we have a second-order ODE and one W\"unschmann condition, for $n = 3$ we have a system of three
second-order PDE's and four W\"unschmann conditions and for $n=4$ we have a system of six second-order PDE's and
ten W\"unschmann conditions.\\

\noindent Summarizing:\newline

a) If we start from a complete integral, $u = Z(x^{a}, s^{i})$ to
the H-J equation, (\ref{HJ3}), then it satisfies the system of
$\frac{n(n-1)}{2}$ second-order PDE's (\ref{PDEIIIc}), with
$\Lambda_{ij}$ satisfying Eqs.\thinspace(\ref{IC}) and the
W\"{u}nschmann-like conditions (\ref{WLC3D'}); In other words, in
the solution space of Eqs.\thinspace(\ref {PDEIIIc}) there is the
naturally defined metric
\begin{equation}
g^{ab} = g^{AB} \theta^a_A \theta^b_B, \label{gab}
\end{equation}
where $g^{AB}$ is given by Eq.\,(\ref{gAB}).

b) If we start with a system of $\frac{n(n-1)}{2}$ second-order
PDE's (\ref{PDEIIIc}), where $\Lambda_{ij}$ satisfy
Eqs.\thinspace(\ref{WLC3D'}) and the integrability conditions,
(\ref{IC}), then in its solution space there exist a natural
$n$-dimensional metric given by Eq.\thinspace(\ref{gAB}). Though
it might appear as if the metric components depend on the
parameters $s^{i}$, the W\"{u}nschmann-like conditions guarantees
that they do not. Furthermore, the solutions $u = Z(x^{a}, s^{i})$
satisfy the H-J equation
\[
g^{ab}\nabla _{a}Z(x^{a}, s^{i})\nabla _{b}Z(x^{a}, s^{i})=1,
\]
with the just determined metric, Eq.\thinspace
(\ref{gab}).\newline

\textbf{Remark 4:} \textit{From the results presented above we conclude that solving the $n$-dimensional H-J
equation, in an $n$-dimensional background space-time, is equivalent to solving a system of $\frac{n(n-1)}{2}$
second-order PDE's.}\newline

In some of the earlier work on the eikonal equation in three- and four-dimensional Lorentzian spaces, it was
proved that the conformal Lorentzian metrics associated with third-order ODE's and pairs of second-order PDE's
satisfying the W\"unschmann condition and generalized W\"unschmann condition, is preserved when the differential
equation is transformed by a contact transformation. For our present case, there is an analogous result given by
the following:\newline

\textbf{Theorem 1:} \textit{Let Eqs.\thinspace (\ref{PDEIIIc}) be
a system of $\frac{n(n-1)}{2}$ second-order PDE's, with}
$\Lambda_{ij}$ \textit{satisfying the conditions (\ref{IC}) and
(\ref{WLC3D'}), and let
\begin{equation}
\overline{\partial}_{ij}\overline{Z} =
\overline{\Lambda}_{ij}(\overline{u}, \overline{w}^i,
\overline{s}^i ),\label{PDEIII*}
\end{equation}
be a second system of\, $\frac{n(n-1)}{2}$\, second-order PDE's
locally equivalent to Eqs.\thinspace(\ref{PDEIIIc}) under the
subset of contact transformations generated by the generating
function
\begin{equation}
H(s, s^{*}, \gamma, u, \bar{s}, \bar{s}^{*}, \bar{\gamma},
\bar{u}) = \bar{u} \mp u - G(s^{i},  \overline{s}^{j}).
\label{SpContact3D}
\end{equation}
Then under this subset of contact transformations the metric given
by Eq.\thinspace (\ref{gab}) is preserved.}\newline

The proof of this theorem is exactly as that presented in
Ref.\thinspace\cite{FKamN} for a system of two second-order PDE's
such that on its space of solutions there is a unique
four-dimensional conformal Lorentzian metric, $g^{ab}$, such that
$g^{ab}u_{,a}u_{,b}=0$. The justification of the form of the
generating function (\ref{SpContact3D}) can be done as in
Refs.\,\cite{GNS1, GNS2, ETG}. For a definition of contact
transformation see Ref.\,\cite{O}  \\

Before closing this section we remark that for, $n = 2$ and  $3$, our general results reduce to that reported in
Refs.\,\cite{GNS1, GNS2, Gallo}. For $n = 4$ we have discovered that in Ref.\,\cite{ETG} is missing one
W\"unschmann condition\, \cite{WLCn4}.\\

\section{The Einstein equations}

We now adopt a new point of view towards geometry on an $n$-dimensional manifold. Instead of a Lorentzian metric
$g^{ab}(x^{a})$ on ${\cal M}$, as the fundamental variable we consider as the basic variables a family of surfaces
on ${\cal M}$ given by $u = constant = Z(x^{a}, s^{i})$ or preferably its second derivatives with respect to
$s^{i}$. From this new point of view these surfaces are basic and the metric is a derived concept. Now we will
find the conditions on $u = Z(x^{a}, s^{i})$ or more accurately on the second order system such that the
$n$-dimensional metric, Eq.\thinspace(\ref{gAB}), be a solution to the Einstein equations.\newline

We start with the Einstein equations in $n$ dimensions, which are
given by (see for example \cite{Myers,Wh}).

\begin{equation}
R_{ab}= 8\pi G \left(T_{ab}-\frac{1}{n-2}g_{ab}T\right),
\end{equation}
with the Ricci tensor given by
\begin{equation}
R_{ab} = \frac{1}{\sqrt{-g}} \frac{\partial}{\partial x^c} (\Gamma^c_{ab}
\sqrt{-g}) - \frac{\partial^2}{\partial x^a \partial x^b} \ln \sqrt{-g}-
\Gamma^c_{ad}\Gamma^d_{bc},  \label{Rab}
\end{equation}
$g = \det(g_{ab})$ and
\begin{equation}
\Gamma^c_{ab} = \frac{1}{2} g^{cd}\left(\frac{\partial
g_{da}}{\partial x^b} + \frac{\partial g_{db}}{\partial x^a} -
\frac{\partial g_{ab}}{\partial x^d} \right),  \label{CS}
\end{equation}
are the Christoffel symbols.\newline

As in the null surface formulation of general relativity, in the present case the Einstein equations are given by

\begin{equation}
R^{ab}Z_{,a}Z_{,b}=8\pi G \left(T^{ab}Z_{,a}Z_{,b}
-\frac{T}{n-2}\right). \label{EE}
\end{equation}

That is, to obtain the Einstein equation in this case, we need to
compute $R^{00}\equiv R^{ab}Z_{,a}Z_{,b}$, which is \textit{one}
of the components of $R^{AB}\equiv R^{ab}\theta _{,a}^{A}\theta
_{,b}^{B}$. From Eq.\thinspace (\ref{gAB}) we have that
$R^{00}=R_{00}$. Using the metric given by
Eq.\thinspace(\ref{gAB}) with coordinates $\theta ^{A}=(\theta
^{0},\theta ^{i})$ in Eq.\thinspace(\ref{Rab}) to compute
$R_{00}$, we find that Eq.\thinspace(\ref{EE}) is equivalent to
\begin{equation}
\frac{1}{2\Delta}\Delta_{,uu}-\frac{1}{2\Delta^2}\Delta^2_{,u}-
\frac{1}{4}\Lambda_{ch,u}\Lambda_{dl,u}g_{hd,u}g_{lc,u}=8\pi G
\left(T^{ab}Z_{,a}Z_{,b} -\frac{T}{n-2}\right),\label{EEF}
\end{equation}
where the covariant components of the metric $g_{hc}=g_{hc}[\Lambda_{mn,u}]$ are obtained as functions of
$\Lambda_{ij}$'s from Eq.\,(\ref{gAB}).\\

At first glance it appears that Eq.\thinspace(\ref{EEF}) cannot be
equivalent to the $\frac{n(n+1)}{2}$ components of the Einstein
equations. However, Eq.\thinspace(\ref{EEF}) is valid for any
value of the $s^i$'s. Thus if we add to Eq.\thinspace(\ref{EEF})
the metricity or W\"{u}nschmann-like conditions, we obtain a set
of consistent equations equivalent to the standard Einstein
equations in $n$ dimensions. The final equations read

\begin{eqnarray}
& & \frac{1}{2\Delta}\Delta_{,uu} -
\frac{1}{2\Delta^2}\Delta^2_{,u} - \frac{1}{4}\Lambda_{ch,u}
\Lambda_{dl,u} g_{hd,u}g_{lc,u} = 8\pi G \left(T^{ab}Z_{,a}Z_{,b}
-\frac{T}{n-2}\right), \nonumber \\
\nonumber\\
& & D_{s^k}[\Lambda_{mn,u}]=\Lambda_{ln,u}\Lambda_{km,w^l} +
\Lambda_{lm,u}\Lambda_{kn,w^l}, \label{Einst***}
\end{eqnarray}
plus the integrability conditions, Eq.(\ref{IC}).\\

As we said in the introduction, we can now view the Einstein equations in either of the two closely related
fashions:

We can consider Eqs.\,(\ref{Einst***}) as $\frac{n(n+1)(n+2)}{6}$
differential equations (of high order) for \textit{the single
function} $Z$. In this case the integrability conditions are not
relevant. Alternatively, the Einstein equations can be considered
as the $\frac{n(n+1)(n+2)}{6}$ equations, Eqs.\,(\ref{Einst***}),
for the $\frac{n(n-1)}{2}$ independent variables $\Lambda_{ij}$.
In this case, the integrability conditions, Eq.\,(\ref{IC}), must
be added but the order of the equations is much lower.

\section*{Conclusions}

In the first part of this work, we have shown that the ideas and procedures developed in our recent
papers\thinspace \cite{GNS1, GNS2,ETG,Gallo}, on the H-J equation can be generalized to the $n$-dimensional H-J
equation on an arbitrary manifold ${\cal M}$. That is, we have shown that on an $n$-dimensional manifold ${\cal
M},$ a definite or indefinite metric, $g_{ab}$, is equivalent to a family of foliations of ${\cal M}$, depending
on ($n-1$) parameters $s^{i}$, described by $u = Z(x^{a}, s^{i})$ that satisfies the W\"{u}nschmann-like
conditions, Eqs.\thinspace(\ref{WLC3D'}). Furthermore, from Eqs.\thinspace(\ref{WLC3D'}) we observe that one can
adopt other points of view, where the $\Lambda_{ij}$ are the basic variables and $u = Z(x^{a}, s^{i})$ is an
auxiliary variable. From this second point of view, Eqs.\thinspace(\ref{WLC3D'}), are simpler but require that we
add the integrability conditions (\ref{IC})
so that a $Z$ does exist.\\

In the second part of this work we have reformulated the Einstein equations in $n$ dimensions as equations for
families of surfaces. If $Z$ is taken as the basic variable then the Einstein equations are equivalent to
Eqs.\thinspace(\ref{EEF}). But if the $\Lambda_{ij}$ are the basic variables the Einstein equations are equivalent
to Eqs.\thinspace(\ref {EEF}) and (\ref{IC}). In both
cases the W\"{u}nschmann equations are needed.\\

To establish our main results we have used a complete integral to the H-J equation on an $n$-dimensional manifold.
We point out that preliminary computations suggest that a similar program can be carry out for the Eikonal
equation in $n$ dimensions. In a future
paper we will present these results.\\

\section*{{Acknowledgments}\newline}

\noindent M.M.M acknowledges the financial support from CONACyT
through a scholarship and G.S.O. acknowledges the financial
support from Sistema Nacional de Investigadores (M\'{e}xico), from
VIEP-BUAP through the grant II 161-04/EXC/G and
from CONACyT through the grant 44515-F. E.G thanks to CONICET for financial support.\\


\begin{thebibliography}{99}
\bibitem{WLCn4} By using the notation if Ref.\cite{ETG} the missing W\"{u}nschmann
condition is given by $$D_s \Psi^*_u = \Psi_u \Lambda^*_w +
 \Psi^*_u \Lambda^*_{w^*} + \Upsilon_u \Lambda^*_R + \Phi_u \Psi_w +
 \Lambda^*_u \Psi_{w^*} + \Psi^*_u \Psi_R.$$
\end{thebibliography}
\end{document}